\documentclass[a4paper,twocolumn,showpacs,amsmath,pra,amssymb,nofootinbib]{revtex4-1} 
\usepackage[T1]{fontenc}
\usepackage[latin9]{inputenc}
\usepackage{times}
\usepackage{color}
\usepackage{xspace}
\usepackage{amssymb,amsmath}
\usepackage{amsbsy}
\usepackage{graphicx}
\usepackage{epstopdf}
\usepackage{float}
\usepackage{hyperref}
\usepackage{adjustbox}
\usepackage{mathrsfs}
\usepackage{bbold}

\newcommand{\eqr}[1]{Eq.~(\ref{#1})}
\newcommand{\eqrs}[1]{Eqs.~(\ref{#1})}
\newcommand{\figr}[1]{Fig.~\ref{#1}}
\newcommand{\figsr}[1]{Figs.~\ref{#1}}
\newcommand{\appr}[1]{Appendix \ref{#1}}
\newcommand{\secr}[1]{Sec.~\ref{#1}}

\newcommand{\braca}[1]{\left(#1\right)}
\newcommand{\bracb}[1]{\left[#1\right]}
\newcommand{\bracc}[1]{\left|#1\right|}

\newcommand{\bracf}[1]{\left\langle #1 \right\rangle}
\newcommand{\bracga}[1]{\langle #1 \rangle}

\newcommand{\bra}[1]{\langle #1|}
\newcommand{\ket}[1]{|#1\rangle}

\newcommand{\diag}{\mathrm{diag}}

\newcommand{\eq}[1]{\begin{equation} #1 \end{equation}}

\newcommand{\balp}{\boldsymbol{\alpha}}

\newcommand{\bsig}{\boldsymbol\sigma}
\newcommand{\bnu}{\boldsymbol\nu}

\newcommand{\bLm}{\boldsymbol\Lambda}
\newcommand{\blm}{\boldsymbol\lambda}

\newcommand{\bA}{\mathbf{A}}
\newcommand{\bB}{\mathbf{B}}
\newcommand{\bC}{\mathbf{C}}
\newcommand{\bD}{\mathbf{D}}
\newcommand{\bE}{\mathbf{E}}
\newcommand{\bF}{\mathbf{F}}

\newcommand{\bM}{\mathbf{M}}
\newcommand{\bR}{\mathbf{R}}
\newcommand{\bQ}{\mathbf{Q}}
\newcommand{\bW}{\mathbf{W}}

\newcommand{\bd}{\mathbf{d}}
\newcommand{\be}{\mathbf{e}}

\newcommand{\Hh}{\hat{H}}
\newcommand{\Nh}{\hat{N}}

\newcommand{\hh}{\hat{h}}
\newcommand{\nh}{\hat{n}}
\newcommand{\ph}{\hat{p}}
\newcommand{\rh}{\hat{r}}
\newcommand{\sh}{\hat{s}}
\newcommand{\xh}{\hat{x}}

\newcommand{\psihdx}{\hat{\psi}^{\dagger}(x)}
\newcommand{\psihx}{\hat{\psi}(x)}

\newcommand{\asc}{a_\mathrm{sc}}
\newcommand{\gid}{g_\mathrm{1D}}

\newcommand{\Dell}{\Delta_\mathrm{L}}

\newcommand{\Efp}{\hat{\bE}^+}
\newcommand{\Efm}{\hat{\bE}^-}

\newcommand{\xph}{\rh_{dl}}
\newcommand{\pph}{\sh_{dl}}

\begin{document}
\title{Manipulation of collective quantum states in Bose-Einstein condensates by continuous imaging}
\author{Andrew~C.~J.~Wade}
\author{Jacob F. Sherson}
\author{Klaus M\o{}lmer}

\affiliation{Department of Physics and Astronomy, University of Aarhus, DK--8000 Aarhus C, Denmark}

\begin{abstract}
We develop a Gaussian state treatment that allows a transparent quantum description of the continuous, nondestructive imaging of and feedback on a Bose-Einstein condensate.
We have previously demonstrated [Phys.~Rev.~Lett.~\textbf{115}, 060401 (2015)] that the measurement back action of stroboscopic imaging leads to selective squeezing and entanglement of quantized density oscillations.
Here, we investigate how the squeezing and entanglement are affected by the finite spatial resolution and geometry of the probe laser beam and of the detector and how they can be optimized.
\end{abstract}

\pacs{03.75.-b, 03.75.Gg, 42.50.Dv, 67.85.-d}

\maketitle


\section{Introduction}

The experimental control of ultracold atomic dynamics has demonstrated landmark achievements through laser and evaporative cooling \cite{ketterle1999a},
engineering of trapping potentials \cite{bloch2008a},
and tuning of atomic interactions \cite{Chin2010a}.
This has motivated the pursuit of quantum technologies with ultracold atoms, such as, quantum computers \cite{ladd2010a}, repeaters \cite{Briegel1998a}, and, simulators \cite{lewenstein2007a,bloch2012a}, and the realization of high precision metrology \cite{Muntinga2013a,Berrada2013a,Burrage2015a}.
As the precision quantum control of ultracold atoms blossoms, new diagnostic and manipulative methods are required.
Nondestructive methods using dispersive light-matter interactions and measurement back action have demonstrated squeezed \cite{Kuzmich2000a,Vasilakis2015a} and entangled \cite{Julsgaard2001a} states, quantum teleportation \cite{Sherson2006a}, and a quantum memory for light \cite{Julsgaard2004a} in room temperature vapour experiments.
In contrast, typical measurements of ultracold atomic systems, such as, imaging after time-of-flight measurements, are destructive and prohibit further interrogation or use of the system. Nondestructive experiments in ultracold atoms have shown great progress
\cite{Andrews1996a,Andrews1997a,Bradley1997a,Stamper1999a,Saba2005a,Sanner2011a,Kaminski2012a,Gajdacz2013a}, and while numerous proposals exist
\cite{Mekhov2012a,Hauke2013a,Hush2013a,Lee2014a,Wade2015a} to manipulate collective quantum states, the exploitation of the quantum nature of the measurement interaction and back action is yet to be realised.

Individual quantum systems undergoing continuous monitoring via a dispersive interaction with a light field (probe) that is subsequently measured, obey stochastic master equations \cite{wiseman2009a}, however, their numerical solution for complex many-body systems is problematic. For the continuous monitoring of a Bose-Einstein Condensate (BEC), stochastic mean-field solutions have been developed \cite{Dalvit2002a,Szigeti2009a}, and full quantum mechanical simulations have been possible for dynamics restricted to a few collective modes \cite{Hiller2012a} while phase-space methods have been employed for larger multimodal treatments \cite{Hush2013a,Lee2014a}.
Provided the atomic system is not significantly perturbed, a Gaussian state treatment of the atomic Bogoliubov excitation modes can be applied around the atomic mean-field, yielding a tractable and transparent description. In our previous work \cite{Wade2015a}, it was thus demonstrated that squeezed and entangled states of atomic modes can be selectively prepared by stroboscopically modulating the probe.
In this article, we use the Gaussian treatment to analyze the measurement back action in detail, including how the finite spatial resolution and geometry of the detector affects the atomic dynamics. In particular, we investigate the interplay between atomic interactions and the measurement back action in the formation of atomic spatial and momentum number correlations.

The paper is structured as follows. In \secr{sec:estab_theory} the Bogoliubov description of the atomic system and the treatment of the light and light-matter interaction are established. In \secr{sec:Gaussian_frame} the Gaussian state formalism and equations of motion for the atomic system are presented. We analyze the evolution of the atomic correlations in \secr{sec:ev_atom_correlations}, while in \secr{sec:first_moments_feedback}, we investigate the application of feedback to counteract the displacements of atomic modes, induced by the measurements.
The production of atomic correlations by spatially inhomogeneous probing and by homogeneous stroboscopic probing is investigated in \secr{sec:x_kx_num_correlations}. \secr{sec:conclusions} summarizes the conclusions and provides an outlook.

\section{Optical probing of Bose Einstein condensate dynamics}\label{sec:estab_theory}
\begin{figure}[h]
\centering
	\includegraphics[width=.3\textwidth]{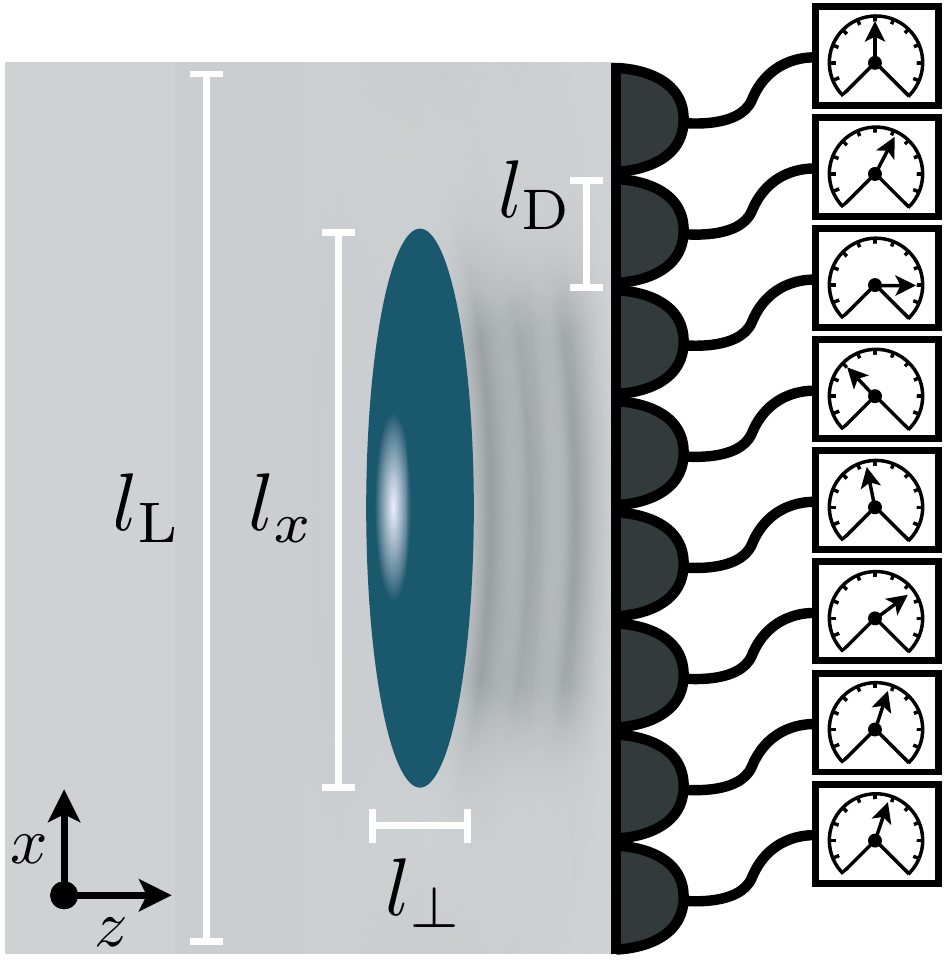}
	\caption{(color online) A far-detuned light field propagating along the $z$-axis interacts dispersively with an atomic ensemble which imprints density-dependent phase shifts across the coherent wavefront. The detection of the optical phase shifts by an array of homodyne detectors (pixels) of widths $l_\mathrm{D}$ reveals information about the atomic density along the BEC axis. The simulated examples are performed for probing of the $D_2$ line ($\sigma_+$ polarised light) of $1000$ $^{87}$Rb atoms, prepared in the $\ket{F,m_F}=\ket{2,2}$ state with axial $\omega_x=2 \pi \times150\mathrm{Hz}$ and radial $\omega_\bot=100\omega_x$ trapping frequencies, and chemical potential $\mu = 2\hbar \omega_x$. \label{fig:system_cartoon}}
\end{figure}

A schematic of the system studied in this article is shown in \figr{fig:system_cartoon}.
As the light interacts with the atomic system, the quantum phase fluctuations across the optical wave front in the light field become entangled with multimodal quantum fluctuations of the atomic density. This causes decoherence of the atomic quantum state, unless one detects the light field and keeps track of the measurement back action on the atomic system.

\subsection{Ultracold bose gases}

We consider harmonically confined 1D ultracold Bose gases with axial (radial) trapping frequency $\omega_{x(\bot)}$ and harmonic length scale $l_{x(\bot)}=\sqrt{\hbar/m\omega_{x(\bot)}}$ (\figr{fig:system_cartoon}), where the tight confinement freezes out the radial motional degrees of freedom \cite{Petrov2004a}. The effective 1D system is described by the second quantized Hamiltonian
\eq{\Hh=\int \psihdx\bracb{H_{\mathrm{1D}} +\frac{\gid}{2}\psihdx\psihx }\psihx dx,\nonumber}
where $H_{\mathrm{1D}}$ is the 1D single atom Hamiltonian with the harmonic potential $V(x)=m\omega_x^2 x^2 /2$, $\psihx$ is the bosonic field annihilation operator, and the 1D interaction strength is given by $\gid=2\hbar^2 \asc/ m l_\bot^2$ with the \textit{s}-wave scattering length $\asc$.
Near zero temperature, the system is well described by a mean field treatment, $\psihx \rightarrow \psi(x) + \delta \psihx$, where the ground state BEC wavefunction $\psi(x)$ (taken to be real) is a solution of the 1D Gross-Pitaevskii equation \cite{Petrov2004a}
\eq{[H_{\mathrm{1D}}+\gid n_0(x)] \psi(x) = \mu\psi(x),\label{eq:GPE}}
and where $\delta \psihx$ denotes the relatively smaller quantum fluctuations about the BEC mean field.
The BEC density is $n_0(x)=\psi^2(x)$, while the chemical potential $\mu$ enforces the BEC population, $N_0=\int\!\! dx n_0(x)$, to the total number of atoms $N$.

The elementary excitations of the BEC, collective center-of-mass, breathing, and higher order modes, are obtained through linearization about the BEC mean field by the Bogoliubov transformation $\delta \psihx=\sum_j [f_j^-(x) \xh_j + i f_j^+(x)\ph_j]$ with dimensionless quadrature observables of the $j$th mode, obeying canonical commutator relations $[\xh_j,\ph_k]=i \delta_{jk}$. With the exception of the zeroth mode, each mode constitutes a quantum harmonic oscillator degree of freedom,
\begin{equation}
\Hh= \frac{\hbar \omega_0}{2} \xh_0^2 +  \sum_{j>0} \frac{\hbar \omega_j}{2} \braca{\xh_j^2 + \ph_j^2},\nonumber
\end{equation}
and their frequencies $\omega_j$ and wavefunctions $f_j^\pm(x)$ are found by solution of the Bogoliubov-de Gennes equations  \cite{Petrov2004a}
\begin{equation}
\bracb{\begin{array}{cc} 0 & \mathcal{L}_+ \\ \mathcal{L}_- & 0 \end{array}}\bracb{\begin{array}{c} f_j^+(x)\\ f_j^-(x) \end{array}} = \hbar \omega_j \bracb{\begin{array}{c} f_j^+(x)\\ f_j^-(x) \end{array}}, \label{eq:BdGs}
\end{equation}
with $\mathcal{L}_\pm=H_{\mathrm{1D}}-\mu + (2\pm1) \gid n_0(x)$ and the normalization $\int \!\!dx f_j^+(x)f_k^-(x)=\delta_{jk}/2$.
The fluctuations of the density about the BEC mean field, $\nh(x,t) = n_0(x)+\nh_\mathrm{nc}(x,t)$,
\begin{multline}
\nh_\mathrm{nc}(x,t)= 2 \psi(x) \sum_{j} f_j^-(x)\xh_j -\sum_j f_j^-(x)f_j^+(x) \\
+{}\sum_{jk}\bracb{f_j^-(x)f_k^-(x)\xh_j\xh_k + f_j^+(x)f_k^+(x)\ph_j\ph_k},\label{eq:nc_density}
\end{multline}
constitute the quantum mechanical property of the atoms that is probed and affected by the optical probing.

The expansion of the field operators around the mean field further yields a pair of functions \cite{Blaizot1986a,Castin1998a}, $[f_0^+(x),f_0^-(x)]=[\psi(x)/\sqrt{2N_0},\sqrt{2N_0} \partial \psi(x)/\partial N_0]$, representing fluctuations of the total atom number and the phase of the BEC mean field. $f_0^-(x)$ thus describes the change of the BEC wavefunction by the addition and removal of BEC atoms, and $\omega_0\equiv 2N_0\frac{\partial \mu}{\partial N_0}$ yields the rate at which superposition states with different BEC atom numbers dephase due to the atomic interactions \cite{Lewenstein1996a,Molmer1998a}. The number conserving Bogoliubov theory \cite{Castin1998a}, used in the stroboscopic schemes presented later, does not include these contributions, but retains the density fluctuations associated with the other Bogoliubov modes in (\ref{eq:nc_density}).

\subsection{Light-matter interaction}

Nondestructive continuous monitoring of the atomic density can be achieved by the dispersive interaction with far-detuned light that is subsequently measured \cite{ketterle1999a,Hammerer2010a}.
For laser detuning $\Dell$ much larger than the excited state hyperfine splitting, the interaction between the light field and a single atom is $\Hh_\mathrm{I} = \Efm \balp\, \Efp$ \cite{Hammerer2010a,Vasilyev2012a} with the positive (negative) frequency component of the electric field $\hat{\bE}^{+(-)}$.
The polarizability tensor operator is
\eq{\balp = -\frac{d_{JJ'}^2}{\hbar \Dell}\bracb{\alpha_0+\alpha_1\hat{F}_z\braca{\be_1\otimes \be_1^* - \be_{-1}\otimes \be_{-1}^*}}, \nonumber}
where $d_{JJ'}=\bracc{\bra{J}|d|\ket{J'}}$ is the reduced dipole matrix element, $\alpha_0$ and $\alpha_1$ are constants depending on the specific atomic structure, $\hat{F}_z$ is the $z$-component of the ground state spin operator, and $\be_q$ are the complex spherical unit vectors. The term proportional to $\alpha_0$ (scalar polarizability) corresponds to an ac Stark shift, while the $\alpha_1$ term (vectorial polarizability) is a spin-state dependent Faraday rotation.

Both the ac Stark shift \cite{Andrews1996a,Andrews1997a} and Faraday rotation \cite{Bradley1997a,Kaminski2012a,Gajdacz2013a} allow spatially resolved imaging of the atomic density, while Faraday rotation also allows characterisation of Bose-Hubbard models \cite{Rogers2014a} and probing \cite{Eckert2007a,Eckert2008a,Bruun2009a,Roscilde2009a,deVega2008a,DeChiara2011,RomeroIsart2012a} and quantum control \cite{Hauke2013a} of spin correlations. The ac Stark shift and Faraday rotation both imprint atomic properties on the phase and polarization of the light field and both must be considered for the characterisation and quantum control of the atomic density. In typical experiments, we have $\alpha_0 \sim \alpha_1$ \cite{Hammerer2010a,Vasilyev2012a}, therefore, measurement of only one will lead to decoherence associated with the atomic entanglement with the other.

In the following, we will assume atoms prepared in the state $\ket{F,m_F}$ and $\sigma_{\pm}$ polarised light, giving the interaction
\eq{\Hh_\mathrm{I} = -\frac{d_{JJ'}^2}{\hbar \Dell} \braca{\alpha_0\pm\alpha_1 m_F} \Efm \!\!\cdot\Efp, \label{eq:sp_int_ham}}
which requires only phase shift measurements for complete information recovery.

\subsection{Treatment of the light field}\label{sec:light_treat}

\begin{figure}[ht]
\centering
	\includegraphics[width=.45\textwidth]{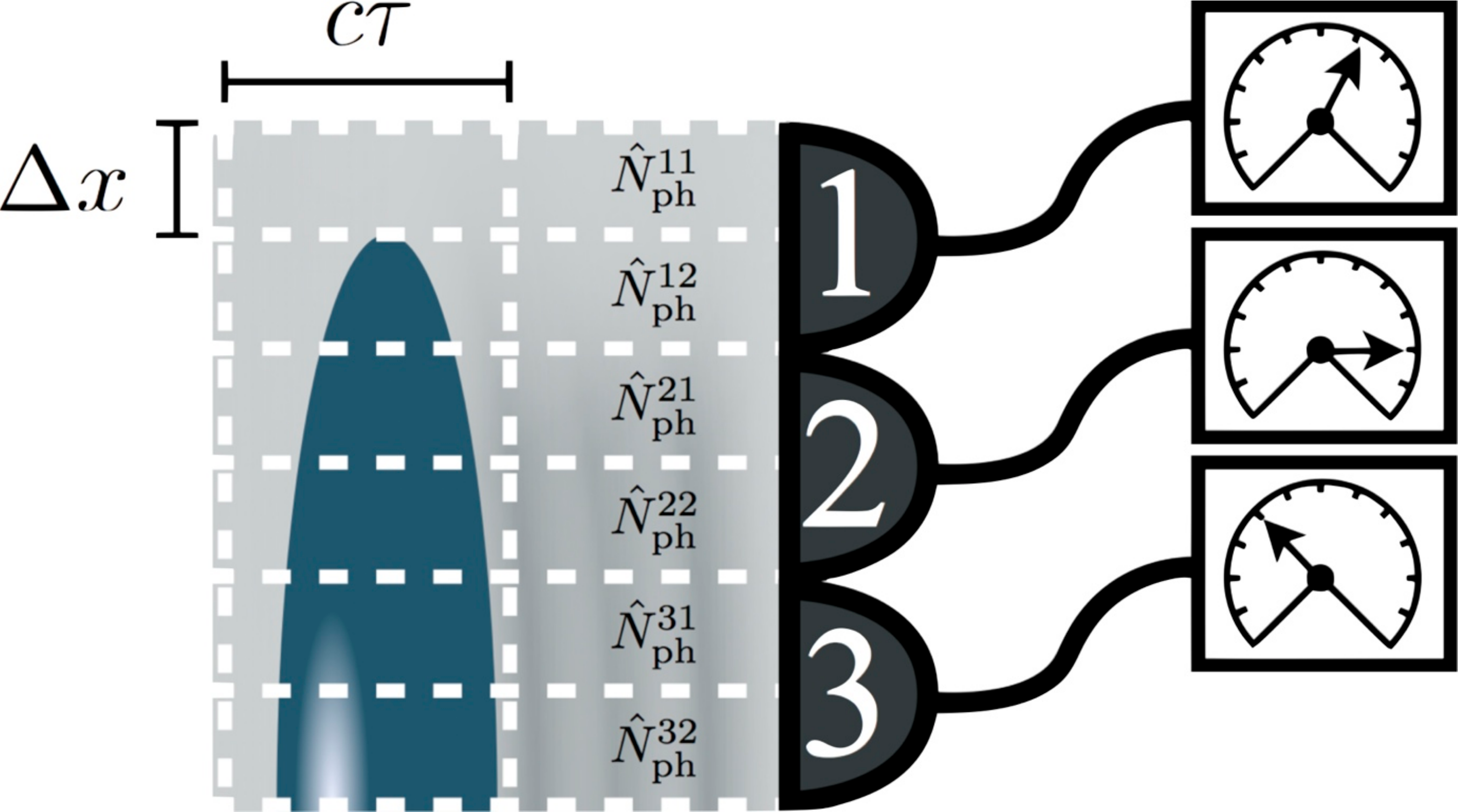}
	\caption{(color online) The planar coherent light field is discretized into cuboidal mode functions for the specific example of $N_\mathrm{L}=2$. During a sequence of time steps of duration $\tau$, sets of light modes along the $x$ direction are detected after interacting with the atomic system. \label{fig:light_discretization}}
\end{figure}

To describe the light-matter interaction and measurement, we develop upon the methodologies of Ref.~\cite{Madsen2004a}, where the planar coherent light field is modelled by discrete cuboidal mode functions (see \figr{fig:light_discretization}) with $\Delta x$, $\Delta y$, and $c\tau$ being the widths in the $x$, $y$, and $z$ directions, respectively, and $c$ is the speed of light.
The photon number operator of each cuboidal mode, labeled $dl$, is
\eq{\Nh^{dl}_\mathrm{ph} \simeq N_\mathrm{ph} + \sqrt{2N_\mathrm{ph}} \xph, \label{eq:num_phase_approx}}
where the average photon number, $\langle \Nh^{dl}_\mathrm{ph} \rangle=N_\mathrm{ph}$, is the same in all modes owing to the planar wavefront assumption.
The number and phase fluctuations, pertaining to initially uncorrelated modes in vacuum, are represented through the dimensionless quadrature observables $\xph\equiv \xph(t)$ and $\pph\equiv \pph(t)$, respectively, with $[\xph,\hat{s}_{d'l'}]=i \delta_{dd'}\delta_{ll'}$.

We assume a temporal discretization such that only single light modes along the $z$ direction interact with the atomic system at any instance, are then immediately detected, and the next set of light modes enter along the $z$ direction. The propagation time of the mode through the BEC and the temporal resolution bandwidth of optical detection are both orders of magnitude faster than the msec-sec timescale of atomic evolution.
Owing to finite spatial resolution, the $d^\mathrm{th}$ detector pixel measures a sum of light quadratures
\eq{\rh_d=\frac{1}{\sqrt{N_\mathrm{L}}}\sum_{l=1}^{N_\mathrm{L}} \xph, \qquad \sh_d=\frac{1}{\sqrt{N_\mathrm{L}}}\sum_{l=1}^{N_\mathrm{L}} \pph, \label{eq:pixel_mode}}
where $N_\mathrm{L}$ is the number of light modes measured by each pixel. This implies a coarse graining of the information retrieved from the light field, and hence, causes a decoherence of the atomic quantum state as will be discussed in \secr{sec:pixel_size}.

Armed with the discrete description (\ref{eq:num_phase_approx}), the dispersive interaction (\ref{eq:sp_int_ham}) with a set of light modes along the BEC axis interacting for time $\tau$, $\Hh_\mathrm{I} =\hbar(\hh_\mathrm{I} + \hh_\mathrm{mf}) (\Delta x/\tau)^\frac{1}{2}$, splits into a BEC mean field contribution,
\eq{\hh_\mathrm{mf} =\kappa \sqrt{\frac{l_\mathrm{L}}{2 N_0}}   \sum_{dl} n_0(x_{dl}) \, \xph,\nonumber}
and a coupling of the quantum fluctuations,
\eq{\hh_\mathrm{I} = \sqrt{2} \sum_{jdl} \kappa_j(x_{dl})  \,  \xh_j \xph, \label{eq:coarse_ham}}
where
\begin{equation}
\kappa_j(x) = \kappa \sqrt{2 l_\mathrm{L}}f^+_0(x)f^-_{j}(x).\nonumber
\end{equation}
We have neglected constant contributions and the relatively smaller $2^\mathrm{nd}$ order terms of the quadrature expansion of $\nh_\mathrm{nc}(x,t)$, c.f., \eqr{eq:nc_density}.
The coupling constant $\kappa = -\braca{\alpha_0\pm\alpha_1 m_F}\sqrt{d_0 \eta}$ \cite{Hammerer2010a} governs the overall measurement strength.
The light field with total photon flux $\Phi$ and cross-sectional area $A_\mathrm{L}$ has the on-resonance optical depth and cross-section, $d_0 = N_0 \sigma_0/A_\mathrm{L}$ and  $\sigma_0 = 3 \lambda^2/2\pi$, respectively. Although the light field is far-detuned, atomic absorption and spontaneous emission occurs at the rate $\eta = \Phi \sigma_0\Gamma^2/4A_\mathrm{L}\Dell^2$ and decoheres the atomic system. This process is neglected in the current treatment, thus, restricted to times shorter than $1/\eta$. Decreasing $\eta$, allowing longer simulation times, requires an increase of the optical depth to maintain constant $\kappa$.

Due to the general coupling $\hh_\mathrm{I}$ of the quantum fluctuations of all the atomic density modes and  light modes,  the detection of the optical phase fluctuations in a pixel $\sh_d$  yields information about a linear combination of many atomic modes. Recalling that the BEC dynamics is frozen in the $y$ and $z$ directions, the discretization in the $x$ (spatial) and $z$ (temporal) directions has consequences for the dynamics, which should, however,  converge as the coarse graining is reduced. In the following section we consider the continuous limits $\Delta x \rightarrow 0$ and $\tau\rightarrow 0$.
In principle, the number and phase replacement (\ref{eq:num_phase_approx}) looses validity as the coherent state amplitudes vanishes in the continuous limits. However, it is the integrated effect of many light modes that is dynamically relevant, and the solution of the ensuing continuous evolution represents the system well.
More crucially, as the light mode volumes shrink, the diffraction of the mode across the finite width of the BEC in the propagation direction ($l_z$) will becomes relevant and is included in the following section.

\section{Establishing the Gaussian Description} \label{sec:Gaussian_frame}

The second-order light-matter interaction (\ref{eq:coarse_ham}) and the quadrature measurement of the light field is compatible with a Gaussian quantum state at all times \cite{Weedbrook2012a}, fully characterised by the first, $[\bR]_j = \bracga{\hat{\bnu}_j}$ and second, $[\bA]_{jk}=\mathrm{cov}(\hat{\bnu}_j,\hat{\bnu}_k)$, moments of the field and atom quadrature variables, $\hat{\bnu}$.
The continuous temporal limit $\tau\rightarrow 0$ yields a description solely in terms of the atomic quadrature variables, $\hat{\bnu} \rightarrow \bracb{\xh_1,\ph_1,\xh_2,\ph_2,\ldots}^T$, (\appr{app:gauss}). The random measurement results of each pixel [simulated through Wiener increments $dW_d(t)$] drives a stochastic evolution of the first moments through
\eq{d\bR=-\bD \bR dt + \bA\bM\, d\bW, \label{eq:MTDOUE}}
where $d\bW=[0,dW_1(t),0,dW_2(t),\ldots]^T$.
The block diagonal matrix with $2\times 2$ blocks
\begin{equation}
[\bD]_{j}  =  \left[ \begin{array}{cc}
	0 & -\omega_j \\
	\omega_j & 0 \\
	\end{array}
	\right], \qquad
	[\bD]_{0}  =  \left[ \begin{array}{cc}
	0 & 0 \\
	\omega_0 & 0 \\
	\end{array}
	\right],\nonumber
\end{equation}
accounts for the harmonic rotation at frequency $\omega_j$ in all $\{\xh_j,\ph_j\}$ phase spaces with the exception of $[\bD]_{0}$ accounting for the spreading of the BEC phase.

The interaction, \eqr{eq:coarse_ham}, couples a linear combination of atomic modes to the light field in each pixel and the measurement back action on each atomic mode $j$ due to the outcome by each pixel $d$ is represented by the rectangular matrix $\bM$, composed of $2\times 2$ blocks
\eq{[\bM]_{jd} = - \frac{2}{\sqrt{l_\mathrm{D}}}\sum_{l}\kappa_j(x_{dl}) \Delta x \left[\begin{array}{cc}
0 & 1 \\
0 & 0 \\
\end{array}\right].\nonumber}
Due to the correlation between atomic modes, the stochastic measurement back action is also correlated among the modes, as formally represented by the atomic covariance matrix $\bA$.

In contrast to the displacement, the covariance matrix (second moments) evolves independently of the measurement results,
\eq{\dot{\bA} =  \bE - \bD \bA  - \bA \bD^T - \bA \bM\bM^T \bA. \label{eq:MRDE}}
The environment back action of the light field is represented by the square matrix of $2\times 2$ blocks
\eq{[\bE]_{jk} =  \sum_{dl} \kappa_{j}(x_{dl})\kappa_{k}(x_{dl}) \Delta x\left[\begin{array}{cc}
0 & 0 \\
0 & 1 \\
\end{array}\right].\nonumber}
In the ideal limit where each light segment is perfectly detected by a single pixel, we have the relation
\begin{equation}
[\bM\bM^T]_{jk}\rightarrow4 \left[\begin{array}{cc}
0 & 1 \\
0 & 0 \\
\end{array}\right] [\bE]_{jk} \left[\begin{array}{cc}
0 & 0 \\
1 & 0 \\
\end{array}\right]. \label{eq:id_M}
\end{equation}

Finally, taking the continuous spatial limit $\Delta x \rightarrow 0$, we have for the environment back action matrix
\begin{equation}
[\bE]_{jk} = \left[\begin{array}{cc}
0 & 0 \\
0 & \kappa^2_{jk} \\
\end{array}\right], \quad \mathrm{with}\quad\kappa^2_{jk} = \int \!\! dx  \kappa_{j}(x)\kappa_{k}(x),\nonumber
\end{equation}
and the measurement back action matrices
\begin{equation}
[\bM]_{jd} = \left[\begin{array}{cc}
0 & \nu_{jd} \\
0 & 0 \\
\end{array}\right], \qquad
[\bM\bM^T]_{jk} = \left[\begin{array}{cc}
K^2_{jk} & 0 \\
0 & 0 \\
\end{array}\right],\nonumber
\end{equation}
with
\eq{\nu_{jd} = -\frac{2}{\sqrt{l_\mathrm{D}}}\int_d \!\! dx  \kappa_{j}(x),\qquad K^2_{jk} = \sum_d\nu_{jd}\nu_{kd}.\nonumber}
For a pixel size ($l_\mathrm{D}$) much smaller than the variations of $\kappa_j(x)$ and $\kappa_k(x')$,  $[\bM\bM^T]_{jk}$ becomes equivalent to the ideal full information retrieval case (\ref{eq:id_M}), i.e., $K^2_{jk} = 4\kappa^2_{jk}$.

\subsection{Optical diffraction: the measurement kernel}\label{sec:m_kernel}

\begin{figure}[ht]
\centering
	\includegraphics[width=.4\textwidth]{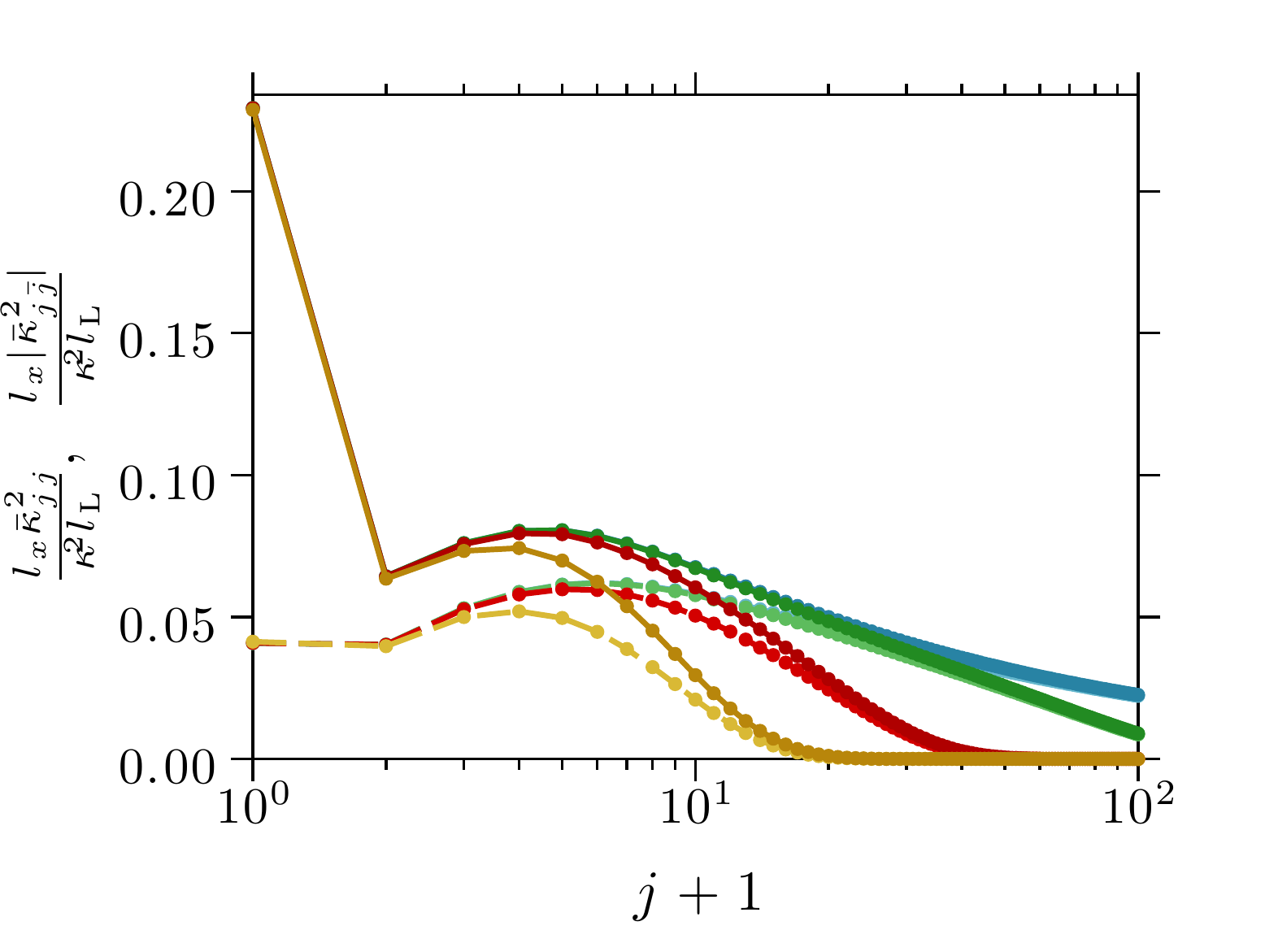}
	\caption{(color online) The diagonal  $\bar{\kappa}^2_{jj}$  (solid) and off-diagonal $\bar{\kappa}^2_{j\bar{j}}$ (dashed) couplings with $\bar{j}=j+2$ are shown for the ideal detector (blue, upper curve) and for $ l_\mathrm{R}=(0.2977, 0.6306, 1.0658)$ [$\lambda = (780.24\mathrm{nm}, 3.5\mu \mathrm{m}, 10\mu \mathrm{m})$] shown from the upper to the lower solid and dashed curves $($green, red, yellow$)$. The lines are to guide the eye between discrete $j$ values. \label{fig:role_of_kernel}}
\end{figure}

The treatments \cite{Dalvit2002a,Szigeti2009a} show that the diffraction of the light field while propagating through the BEC leads to a resolution limit (Rayleigh length) $l_\mathrm{R} = (l_\bot \lambda)^\frac{1}{2}$, representing the smallest width of a light mode that has a limited diffraction across the BEC. Crudely, tighter light modes will diffract more across the BEC and hence interact with a broad range of spatial regions.
To account for diffraction, and obtain agreement with \cite{Dalvit2002a,Szigeti2009a} for $l_\bot > \lambda$, we include the measurement kernel $\mathcal{K}_\alpha \braca{x}$,
\begin{equation}
\mathcal{K}_\alpha \braca{x} = \frac{1}{2\pi }\int \!\! dk e^{-\frac{( \alpha  l_\mathrm{R} k)^4}{64\pi^2}  } e^{i k x} . \label{eq:m_kernel}
\end{equation}
The environment back action couplings become
\begin{equation}
\kappa^2_{jk}\rightarrow \bar{\kappa}^2_{jk} = \int \!\! dx \!\! \int \!\! dx' \mathcal{K}_{\sqrt[4]{2}} (x-x') \kappa_{j}(x)\kappa_{k}(x'),\nonumber
\end{equation}
while the measurement back action couplings become
\begin{equation}
\nu_{jd} \rightarrow \bar{\nu}_{jd} = -\frac{2}{\sqrt{l_\mathrm{D}}}\int_d \!\! dx \!\! \int \!\! dx'  \mathcal{K}_{1} (x-x')\kappa_{j}(x').\nonumber
\end{equation}
The diffraction kernel smears out spatial features smaller than $l_\mathrm{R}$ and thus decouples highly excited atomic modes with spatial variation scaling as $\sim l_x/j$ (\figr{fig:role_of_kernel}). Conversely, by reducing $l_\mathrm{R}$ one increases the coupling to higher modes and thus heats the atoms as observed in the simulation of cooling of a BEC via probing and feedback \cite{Szigeti2009a,Szigeti2010a,Hush2013a}.

\section{Evolution of atomic correlations} \label{sec:ev_atom_correlations}

Having specified the theoretical formalism, we now turn to the dynamics of the probed atomic system. The covariance matrix describes correlations of the system (e.g., squeezing, purity, entanglement) and evolves independently (\ref{eq:MRDE}) of the stochastic measurement results. We shall first analyse this evolution, and later in \secr{sec:first_moments_feedback}, we shall return to the stochastic evolution of the displacements (\ref{eq:MTDOUE}) and the application of feedback.

\subsection{Evolution under the continuous measurement} \label{sec:temporal_res}

\begin{figure}[h]
\centering
	\includegraphics[width=.4\textwidth]{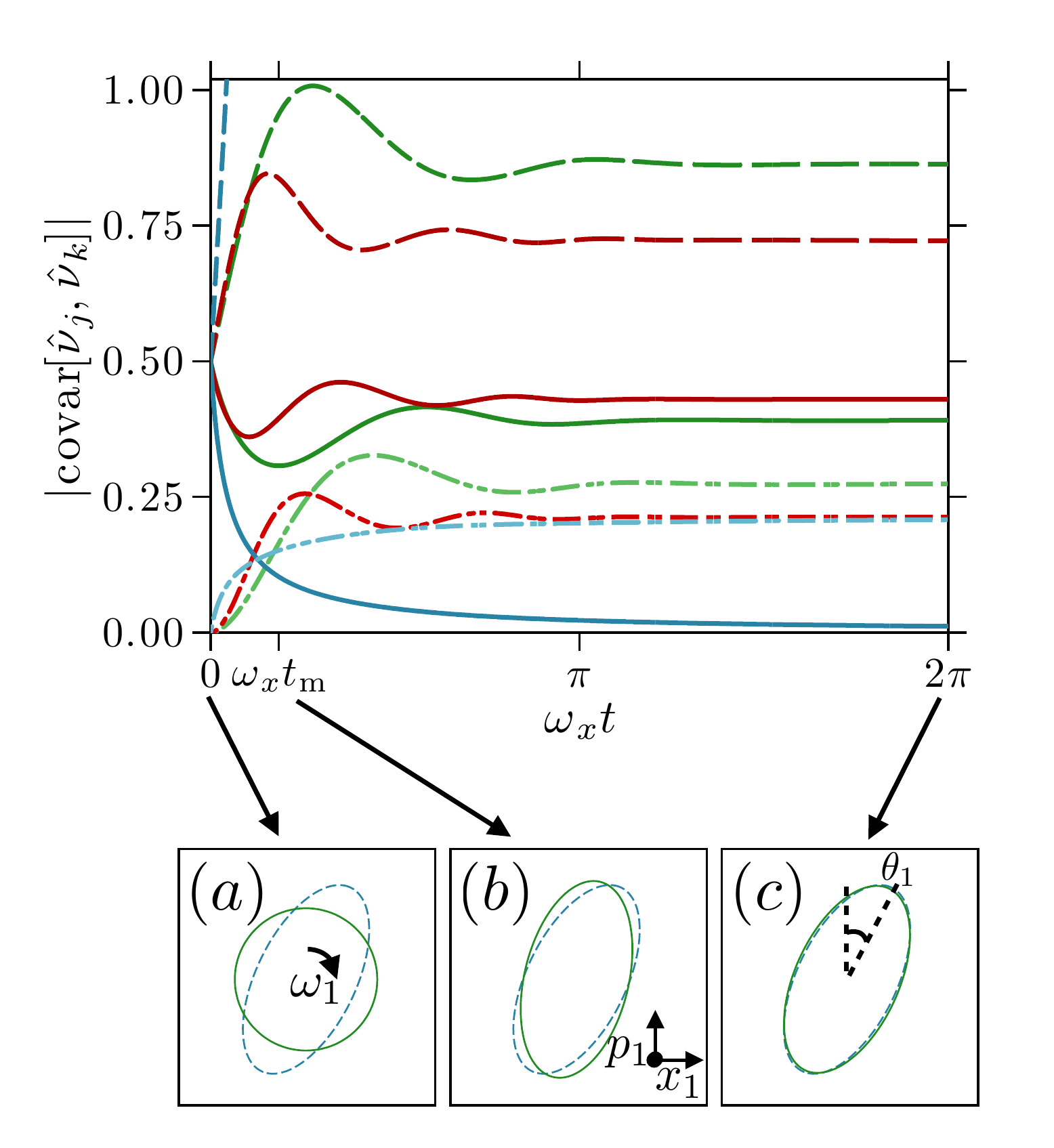}
	\caption{(color online) The temporal dynamics of $\mathrm{var}\,[\hat{x}_j]$ (solid), $\mathrm{var}\,[\hat{p}_j]$ (dashed), and $\mathrm{covar}\,[\hat{x}_j,\hat{p}_j]$ (dashed-dotted), is shown for the zeroth (blue), first (green), and second (red) modes, with $\bar{\kappa}^2_{11}=\omega_x$.
In the lower panels, the uncertainty ellipse of the Gaussian Wigner function of the first mode (green) is illustrated at times $\omega_x t=(0,\omega_x t_\mathrm{m},2\pi)$ with the corresponding steady-state prediction \eqr{eq:A_SS} (dashed blue).
Initially a circle (a), the distribution is squeezed along the $x_1$-axis. However, the antisqueezed $p_1$ quadrature rotates $x_1$, yielding a transient minimum squeezing of $\hat{x}_1$ (b). Ultimately, the dynamics average yielding a steady-state (c).
\label{fig:vars}}
\end{figure}

The measurement strengths $\bar{\kappa}^2_{jj}$ must be comparable to the excitation frequencies of the atomic system $\{\omega_j\}$, to affect the quantum states of the collective atomic density oscillations.
This is reflected in \figr{fig:vars} illustrating the variances $\mathrm{var}\,[\hat{x}_j]$ and $\mathrm{var}\,[\hat{p}_j]$ and covariances $\mathrm{covar}\,[\hat{x}_j,\hat{p}_j]$, assuming a perfectly resolving detector (\ref{eq:id_M}). Figure \ref{fig:vars} demonstrates that density oscillation modes ($j>0$) feature a transient minimum squeezing of the measured $\hat{x}_j$ quadrature before the anti-squeezed $\hat{p}_j$ rotates into $\hat{x}_j$. Thereafter, a steady-state is reached depending on the ratio of the measurement rate to the oscillation frequencies.
The steady-state covariance matrix of the $j$th harmonic mode $\bA^\mathrm{ss}_{jj}$ is predicted quite well by assuming the modes are probed independently, i.e., $\bar{\kappa}^2_{jk} \rightarrow 0$ for $j \neq k$, which yields
\begin{equation}
\bA^\mathrm{ss}_{jj}=\frac{1}{4\tilde{\kappa}_{\omega_j}}\bracb{\begin{array}{cc} [2(a_j-1)]^\frac{1}{2} & a_j-1\\ a_j-1 & [2(a_j-1) a_j^2]^\frac{1}{2}  \end{array}},\label{eq:A_SS}
\end{equation}
where $\tilde{\kappa}_{\omega_j}\equiv \bar{\kappa}^2_{jj}/\omega_j$ and $a_j=[1+4\tilde{\kappa}_{\omega_j}^2]^\frac{1}{2}$.
The squeezing of $\hat{x}_j$ becomes significant for $\tilde{\kappa}_{\omega_j} \gg 1$, while for $1\gg \tilde{\kappa}_{\omega_j}$, $\bA^\mathrm{ss}_{jj}$ is only slightly perturbed from initial vacuum fluctuations. Also owing to the competition between the harmonic rotation and the squeezing of $\xh_j$ by the measurement back action, maximal squeezing occurs along the quadrature $\hat{q}_j=\hat{x}_j\cos\theta_j-\hat{p}_j\sin\theta_j$ [\figr{fig:vars}(c)] where $\theta_j=1/(4\tilde{\kappa}_{\omega_j})^\frac{1}{2}$ and $\theta_j=\pi/4-\tilde{\kappa}_{\omega_j}/2$ in the two regimes, respectively.

Via stroboscopic probing we can perform selective QND-like squeezing and entanglement of quadrature variables associated with the Bogoliubov mode density oscillations \cite{Wade2015a}. For example, a squeezed state of the $j^\mathrm{th}$ mode with $\xh_j(t)=\xh_j^0 \cos \omega_j t+\ph_j^0 \sin \omega_j t$ is produced with squeezed $\xh_j^0$ if the system is probed by a train of pulses, timed such that $\xh_j(t)\propto \hat{x}_j^0$ to avoid the anti-squeezing at intermediate times.
The production of a squeezed oscillator by nondemolition stroboscopic probing was initially proposed in the context of gravity wave detection \cite{Braginskii1978a,Thorne1978a,Braginskii1980a}, and, has found application in mechanical oscillators \cite{Ruskov2005a,Suh2014a} and oscillator spin states of atomic ensembles \cite{Vasilakis2011a,Vasilakis2015a}, in which, it was recently demonstrated \cite{Vasilakis2015a}. Although, the BEC is not a single mode system, provided there is no degeneracy or rational frequency ratio between coupled modes \cite{Wade2015a} the squeezing is mode selective. In \secr{sec:strobo_den_fluct}, we will thus analyze the use of stroboscopic probing to squeeze specific density oscillations.

In addition to the individual modal dynamics, intermodal correlations, and potential entanglement, develop via the off-diagonal couplings $\bar{\kappa}^2_{jk}$, before succumbing to a steady-state for continuous probing. In analogy to the stroboscopic squeezing, entanglement between two modes is established by squeezing their collective variables.
The role of the detector geometry in the generation of entanglement by stroboscopic probing will be investigated in the following section.

\subsection{The role of detector resolution}\label{sec:pixel_size}

\begin{figure}[h]
\centering
	\includegraphics[width=.48\textwidth]{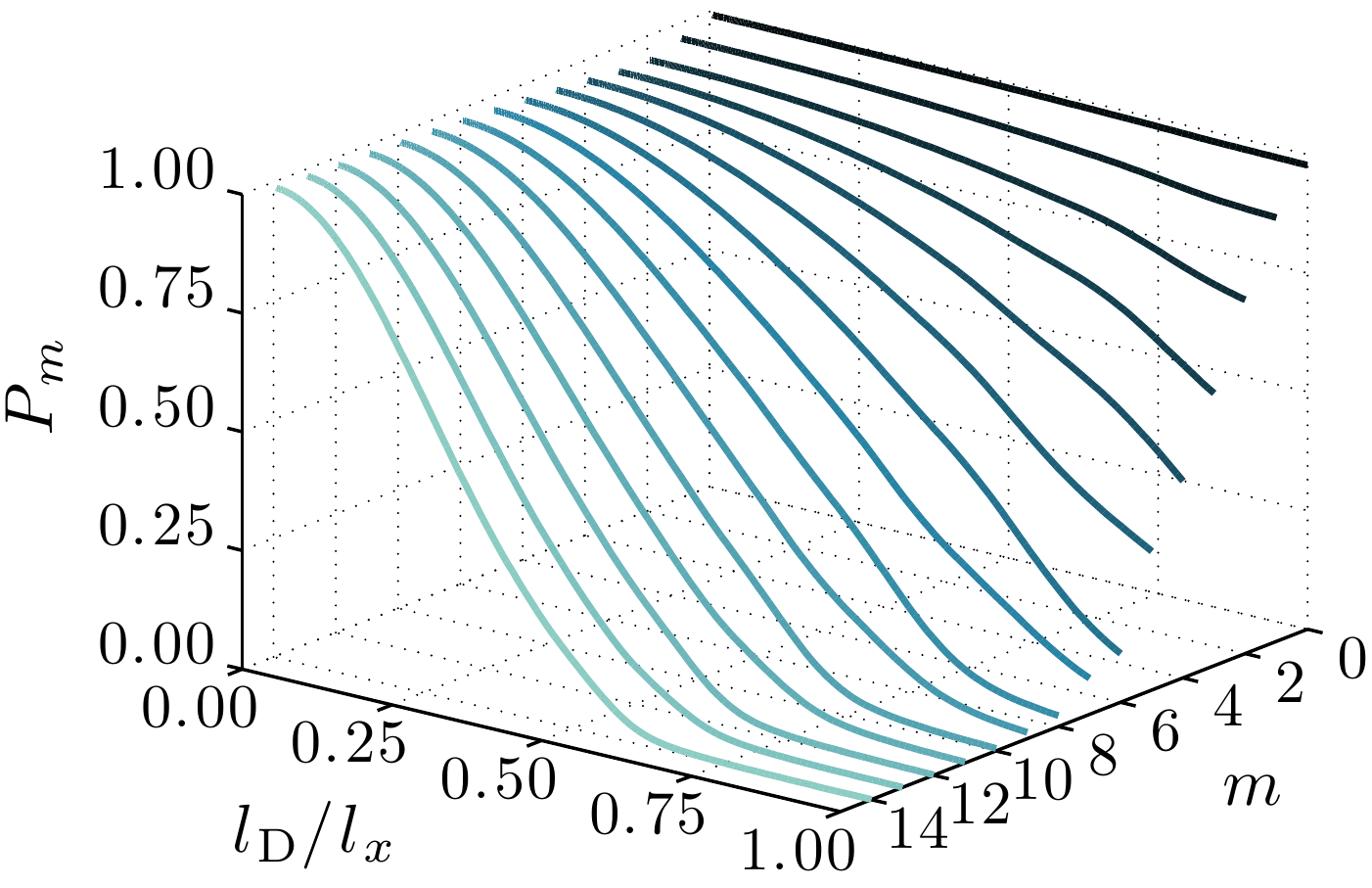}
	\caption{ (color online) The purity $P_m$ of the subset of the first $m+1$ modes is shown as a function of detector resolution $l_\mathrm{D}$ at time $\omega_x t = \pi$ for the simulation of \figr{fig:vars} where the majority of the density oscillations have reached, or are close too, steady-state covariances.  \label{fig:pixel_res}}
\end{figure}

The previous section presented results for perfect spatial detection of the light field.
Accounting for finite detector resolution amounts to tracing-out light field modes outside of the associated spatial bandwidth. Consequently, the measurement cannot restore full coherence to the atomic system.
This also accounts for unresolved spatial structures of the light field within individual pixels.
The associated decoherence is demonstrated in \figr{fig:pixel_res}, quantified by the purity, $P_m=\mathrm{Tr} (\hat{\rho}_m^2)=1/2^m \mathrm{det}(\bA_m)^{\frac{1}{2}}$, of the subsystem up to the $m$th mode [$\hat{\rho}_m=\mathrm{Tr}_{n>m}(\hat{\rho})$] with the covariance matrix $\bA_m$.
As $m$ is increased, $P_m$ becomes increasingly sensitive to $l_\mathrm{D}$ owing to the more rapid spatial variations (scale as $l_x/j$) of the higher-order modes included.
We note that even in the case of the ideal detector, $P_m \leq 1$, due to the decohering effect of entanglement with other atomic modes established by the probing.

\begin{figure}[h]
\centering
	\includegraphics[width=.48\textwidth]{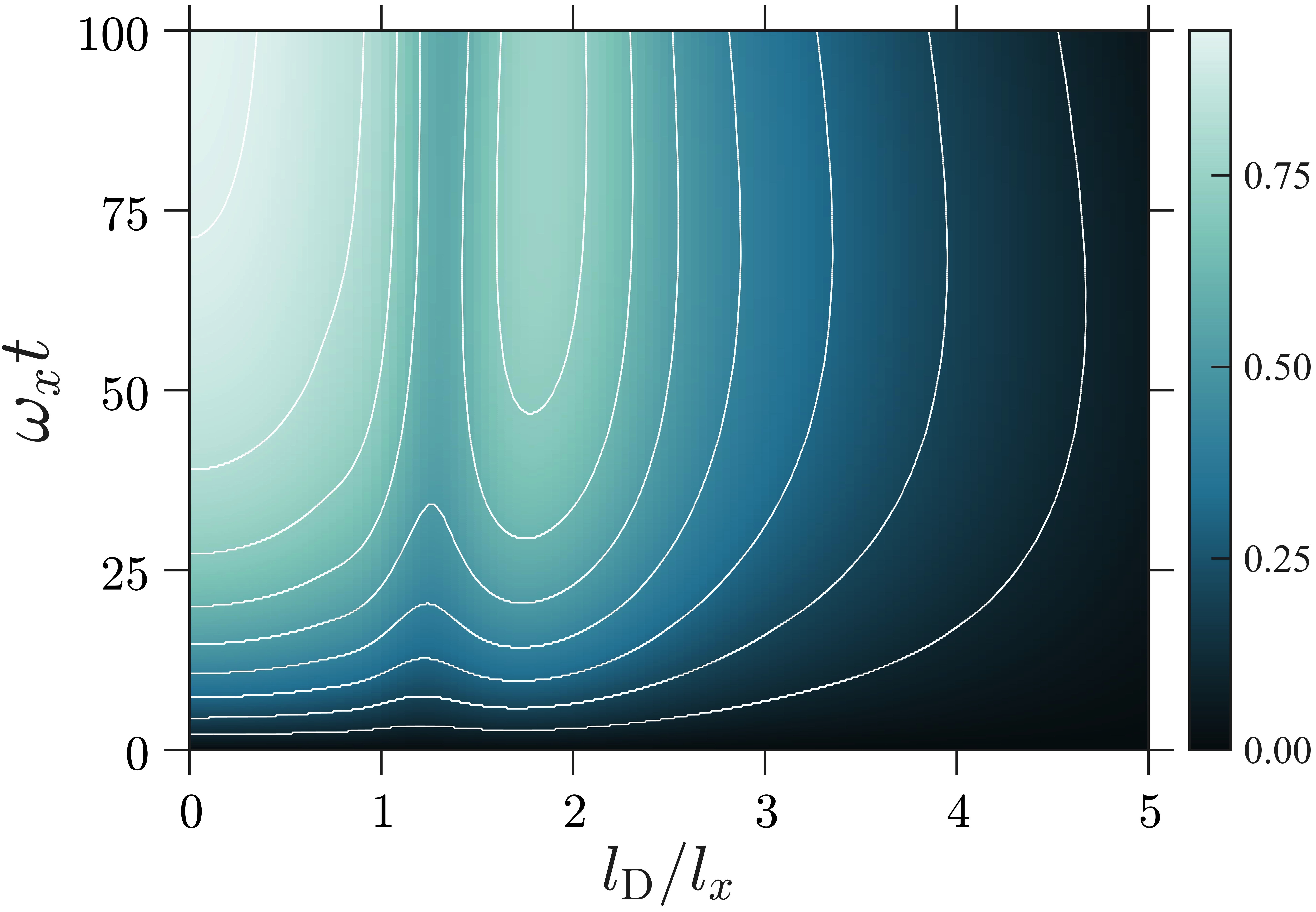}
	\caption{ (color online)
The bipartite entanglement, $E_{13}$, between modes $1$ and $3$, obtained by stroboscopic probing at the sum frequency $\varpi=\omega_1+\omega_3$, is shown as function of pixel resolution and probing time. The simulation assumes probe pulses with $\kappa^2 = 30\omega_x/2\pi$, centred on times $\varpi t_l=[0, 2\pi, 4\pi,\ldots,2(n-1)\pi]$ with identical durations $\varpi \Delta t= 2\pi\times 0.03$. The degree of entanglement is represented by the color brightness, and the white contours are to guide the eye. \label{fig:ENT}}
\end{figure}

The loss of information can also limit the amount of squeezing or entanglement of the atomic modes. Indeed, a completely unresolved density oscillation, $j$, pertains to a thermal state with thermal population $\bar{\kappa}^2_{jj} t/2$ in the long-time limit.
We consider the effect on the entanglement generation between modes $1$ and $3$ for the stroboscopic entanglement generation protocol of \cite{Wade2015a} in \figr{fig:ENT}.
Entanglement may form between the atomic modes because the modal signatures embedded by the light-matter interaction on the light field are indistinguishable. Temporal distinguishability due to different oscillation frequencies is avoided by stroboscopically probing at the sum frequency $\varpi=\omega_1+\omega_3$, while partial spatial distinguishability remains the limiting factor.
The bipartite entanglement is quantified by the logarithmic negativity $E_{13}=\log_2\bracc{\bracc{\hat{\rho}_{13}^\mathrm{Tp}}}_\mathrm{Tr}$ \cite{Weedbrook2012a} of the reduced density matrix  of modes $1$ and $3$ $\hat{\rho}_{13}$, where $\mathrm{Tp}$ denotes the partial transpose and $\bracc{\bracc{\cdot}}_\mathrm{Tr}$ the trace norm. The effective QND probing result converges to the asymptotic limit $E^{13}_{\mathrm{QND}} = \log_4[\frac{1+\beta_{13}}{1-\beta_{13}}]$, where the spatial distinguishability of the mode is parameterised through the spatial mode functions by $\beta_{jk} = |\bar{\kappa}^2_{jk}|/[\bar{\kappa}^2_{jj}\bar{\kappa}^2_{kk}]^{\frac{1}{2}}$.

In \figr{fig:ENT}, the growth of the entanglement is most rapid, and attains the maximal value, for the ideal detector ($l_\mathrm{D}=0$). However, similar entanglement is shown to be generated over a broad range of pixel widths $l_\mathrm{D}$.
Ideally, the pixel geometry retrieves all the information imprinted on the light field about modes $1$ and $3$, while, owing to the selective nature of the stroboscopic probing, correlations with the rest of the atomic system are negligible. Here, the effectiveness of the information retrieval is represented by $K^2_{jk} = \sum_d\nu_{jd}\nu_{kd}$, where the ideal case (\ref{eq:id_M}) is $K^2_{jk} = 4\bar{\kappa}^2_{jk}$.
For a given detector resolution, the pixel geometry can be used to optimize $K^2_{jk}$. Consider the large pixel regime, $l_\mathrm{D} \geq l_\mathrm{x}$.
If the pixel $d=0$ is centred with the BEC, even modes are primarily measured, as the total field it detects is insensitive to the odd modes, i.e., $\bar{\nu}_{j0}\sim \int_{-l_\mathrm{D}/2}^{l_\mathrm{D}/2} \! f^+_0(x)f^-_{j}(x)dx \sim 0$ for odd $j$.
Conversely, if a pixel boundary is centred with the BEC, odd modes are primarily measured owing to the orthogonality with the zeroth mode [$\int_0^\infty \! f^+_0(x)f^-_{j}(x)dx = 0$ for even $j$]. The latter is the geometry considered here, and, although the odd modes $1$ and $3$ feature entanglement, the even modes are decohered for most of the pixel range in \figr{fig:ENT}.
The second maxima of the entanglement along $l_\mathrm{D} = 2l_x$ is due to an interplay between pixel geometry and modal structure, yielding a more optimal geometry.

\subsection{The role of atomic interactions}\label{sec:role_of_int}

The atomic interactions change the properties, and in particular the frequency spectrum, of the density oscillations and, hence, the dynamical response of the system to the probing. From the noninteracting to the interaction dominated Thomas-Fermi regime, the excitations go from the single particle excitations with the harmonic spectrum $\omega_j = \omega_x j$, to the collective excitations with the irregular spectrum $\omega_j = \omega_x\sqrt{j(j+1)/2}$ \cite{Stringari1998a,Ho1999a}.

In the noninteracting case, some integrals involving Hermite polynomials \cite{Titchmarsh1948a,Busbridge1948a} yield
\eq{ \kappa^2_{jk} = \begin{cases} \frac{\kappa^2 l_\mathrm{L} i^{j-k}}{\pi\sqrt{2j!k!}l_x} \Gamma \! \braca{\frac{j+k+1}{2}}&\mbox{if } j+k \mbox{ is even}.\\0 & \mathrm{otherwise,} \end{cases}\nonumber}

For the Thomas-Fermi regime, analytic solutions for $f^\pm_j(x)$ \cite{Petrov2000a} yield ($\kappa^2_{j0} = \kappa^2_{0j}$)
\eq{\kappa^2_{jk} =\begin{cases} \frac{\kappa^2 l_\mathrm{L}}{4l_x}\sqrt{\frac{2\omega_x}{3\omega_0}} \delta_{j0} &\mbox{if } k=0,\\ \frac{\kappa^2 l_\mathrm{L}\hbar\omega_j }{4N_0 g_\mathrm{1D}} \delta_{jk} & \mathrm{otherwise} \end{cases}\nonumber}
illustrating that, in general, the measurement couplings $\bar{\kappa}^2_{jk}$ decrease with increasing interactions.
The dominant density variation occurs as an interference between the BEC mean
 field and the Bogoliubov mode, \eqr{eq:nc_density}, yielding an enhancement of the signal of the mode quadrature by $\sqrt{N_0}$. With increasing interactions, however, the overlap of these modes is reduced, as overlap with the BEC mean field is energetically unfavourable. The signal gradually becomes comparable to the small second-order terms in the density [\eqr{eq:nc_density}] that where neglected in the light-matter interaction (\secr{sec:light_treat}). As our stroboscopic probing only works in the absence of degeneracies in the excitation spectra, we must, however, incorporate interactions. By choosing parameters, $Ng_\mathrm{1D}/\hbar \omega_x l_x=4.953$, leading to $\mu =2\hbar \omega_x$, we obtain a suitable compromise with a reasonable coupling and a sufficiently irregular spectrum for low energy excitations.

\section{Stochastic evolution of displacements and feedback} \label{sec:first_moments_feedback}

So far, we considered only the atomic covariance matrix, \eqr{eq:MRDE}, while we know that the measurement back action also leads to stochastic displacements of the Bogoliubov mode quadratures, cf., \eqr{eq:MTDOUE}. When observed, these displacements are coherent and   correspond to a modification of the Gross-Pitaevskii wavefunction about which the atomic system is linearized. We shall now address the stochastic evolution and discuss how one may employ feedback to maintain the ground state Gross-Pitaevskii wavefunction, \eqr{eq:GPE}.

The application of feedback to maintain a BEC ground state has been explored with phase-space methods \cite{Hush2013a}, where a steady-state close to the ground state is achieved, provided sufficient control of the shape of the feedback potential. Here, the optimal feedback protocol of a mode is readily identifiable.
Damping of the deterministic evolution of the $j$th mode is achieved through mode matching, and feedback of the measured $\langle \hat{p}_j (t)\rangle$, via a single atom position dependent Hamiltonian $H_\mathrm{f} = \hbar \sum_k h_k(x) \langle \hat{p}_k (t)\rangle$, where $h_k(x)=2\epsilon\omega_j f_j^+(x) \delta_{jk} /\sqrt{n_0(x)}$, i.e., an adjustable potential.
In the dynamics, this potential yields the replacement of the $j$th block of $\bD$ in the evolution of the displacements \eqr{eq:MTDOUE} by
\begin{equation}
[\bD]_{j}=\left[\begin{array}{cc} 0 & -\omega_j \\ \omega_j & 2\epsilon \omega_j \\ \end{array}  \right].\label{eq:damping_D}
\end{equation}
The first and second density oscillation modes can be addressed by varying the trap position and strength, and the control of the higher-order modes [$j$th-order (Hermite) Legendre polynomials in the (non)interacting case] may be provided by adaptive lightshift potentials generated through a micro-mirror array \cite{Goorden2014a}, a spatial light modulator \cite{McGloin2003a}, or, an acousto-optic deflector \cite{Ryu2013a}.

\begin{figure}[h]
\centering
	\includegraphics[width=.45\textwidth]{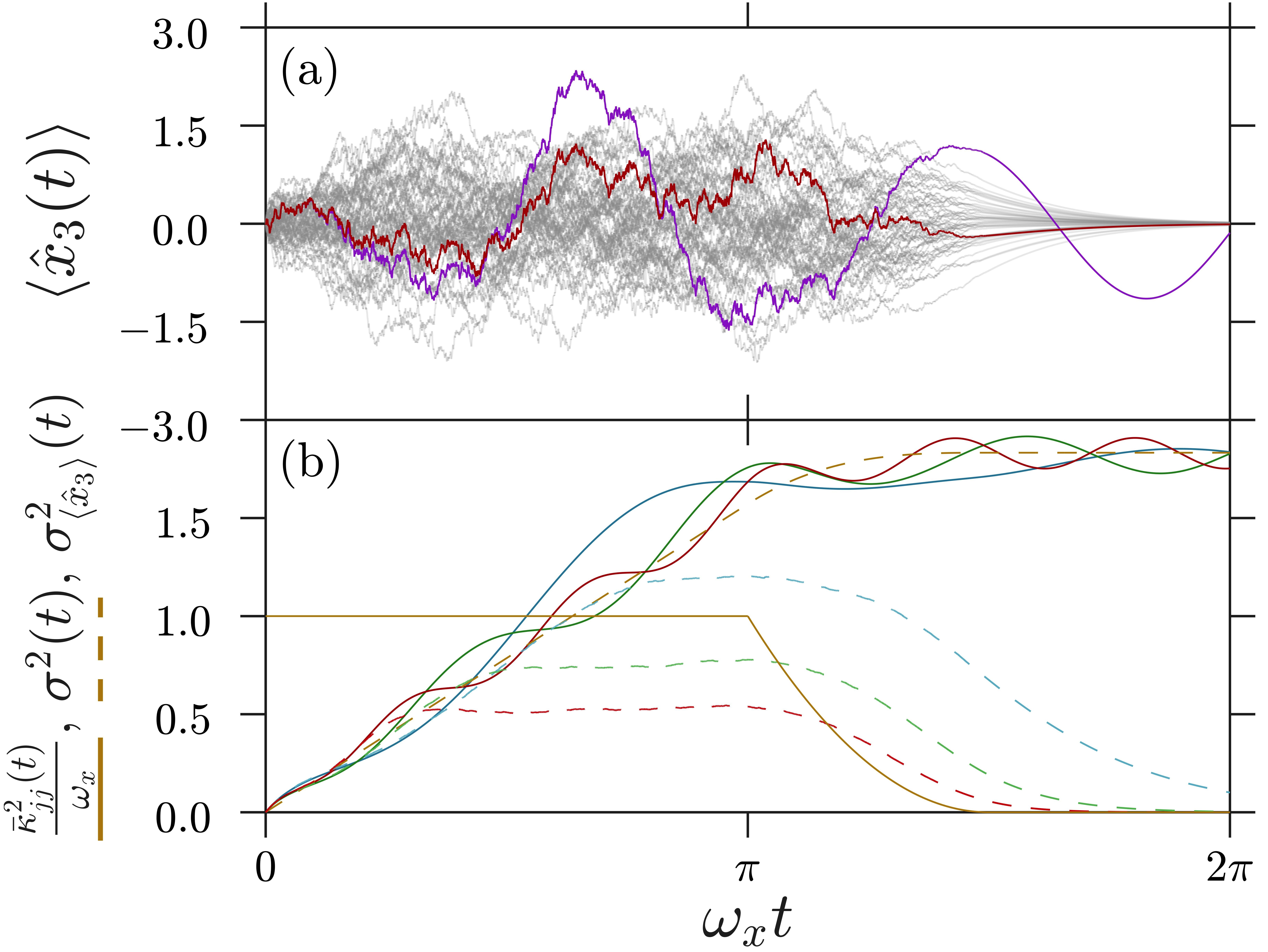}
	\caption{(color online) (a) The grey curves show individual trajectory results for $\bracf{\hat{x}_3(t)}$. The purple (red) curve illustrates a single, undamped (damped) trajectory with the same particular measurement record.    The measurement strength is ramped down after $\omega_x t =\pi$ as illustrated (solid-gold) by $\bar{\kappa}^2_{jj}(t)$ in (b) with $j=3$. (b) also shows the variances, $\sigma^2_{\langle\hat{x}_j\rangle}(t)$, of $10^4$ trajectories for the (un)damped cases of $j=1$-$3$ by the (solid)dashed blue, green, and red lines, respectively. The undamped $\sigma^2_{\langle\hat{x}_j\rangle}(t)$ oscillates around $\sigma^2(t) = \int_0^t \! dt' \bar{\kappa}^2_{jj}(t')/2$ (dashed-gold).  \label{fig:traj_ex}}
\end{figure}

An example of an undamped trajectory is illustrated in \figr{fig:traj_ex}(a, purple) with constant measurement strength until $\omega_x t = \pi$, thereafter, ramped to zero as illustrated in \figr{fig:traj_ex}(b, solid-gold).
The random measurement results cause diffusive harmonic motion, progressively giving way to deterministic oscillatory motion while the measurement strength is ramped to zero. For the same measurement realization, the damped trajectory ($\epsilon =1$) is shown \figr{fig:traj_ex}(a, red), where the excursion of the diffusion is reduced and is critically damped as the measurement weakens.

The (un)damped standard deviations of the trajectories are shown in \figr{fig:traj_ex}(b) for $j=1$-$3$ by the (solid)dashed-lines. Without damping, each case oscillates around $\sigma^2(t) = \int_0^t \! dt' \bar{\kappa}^2_{jj}(t')/2$ (dashed-gold), while with damping, a steady-state is approached during the continuous measurement phase, $0\leq \omega_x t \leq \pi$. The differences between these steady-states are due to the damping occurring at different rates $\omega_j$.
Using \eqr{eq:A_SS}, the steady-state $\bA^{\mathrm{ss}}_\mathrm{ens}$ can be predicted,
and upon minimising the ensemble average of the energy $\hbar \omega_j(\langle\hat{x}_j\rangle^2+\langle\hat{p}_j\rangle^2)/2$, w.r.t.~$\epsilon$, the minimized energy is $2\bar{\kappa}^2_{jj}$ with $\epsilon =1$, and,
\begin{equation}
\bA^{\mathrm{ss}}_\mathrm{ens}=\tilde{\kappa}_{\omega_j} \left[\begin{array}{cc} 3 & -1 \\ -1 & 1 \\ \end{array}  \right],\nonumber
\end{equation}
for weak probing, $\tilde{\kappa}_{\omega_j} \ll 1$.
In the case of strong probing, $\tilde{\kappa}_{\omega_j} \gg 1$, the minimized energy is $3\bar{\kappa}^2_{jj}$ with $\epsilon =(1+4\tilde{\kappa}_{\omega_j})^\frac{1}{2}/2$, and,
\begin{equation}\label{eq:damped_A_ens}
\bA^{\mathrm{ss}}_\mathrm{ens}=  \tilde{\kappa}_{\omega_j} \left[\begin{array}{cc}  5 & - \tilde{\kappa}_{\omega_j}^{-\frac{1}{2}} \\ -\tilde{\kappa}_{\omega_j}^{-\frac{1}{2}} & 1 \\ \end{array}  \right].
\end{equation}
For weak probing, the deterministic motion dominates, and hence, the regime of critical damping ($\epsilon =1$) of the deterministic part of the evolution is optimal. While, for strong probing, the overdamped ($\epsilon >1$) regime is more efficient. The different standard deviations of the quadratures are due to the specific damping of the $p$ quadrature, c.f., \eqr{eq:damping_D}. As $\tilde{\kappa}_{\omega_j}$ increases, the relative importance of the correlations between the quadratures decreases [c.f., \eqr{eq:damped_A_ens}], as the (larger) random displacements due to the measurement back action dephase the individual trajectories.

For all examples presented in this article, with the exception of \figr{fig:vars} at times $\omega_x t > \pi$, the coherent displacements correspond to up to $10\%$ population outside of the BEC mode. This can be reduced through different trapping geometries where the measurement kernel [\eqr{eq:m_kernel}] reduces the number of addressed modes or through more elaborate feedback schemes. In the case of the damping presented here, the targeting of many modes ($10$s-$100$s) can lead to the feedback playing a small (stochastic) role in the dynamical evolution of the covariance matrix through the second-order terms in the density (\ref{eq:nc_density}). As the measurement strength is reduced, the stochastic role of the measurement back action weakens, \figsr{fig:traj_ex}. When the probing is completely absent, the excitation associated with the coherent displacements of the modes (first moments) can be critically damped ($\epsilon =1$), without disruption of the covariance matrix (second moments). In the case of stroboscopic probing, the ideal critical damping can be performed on the targeted mode bringing the first moments close to zero for the entire course of the state preparation \cite{Wade2015a}.

\section{Spatial and momentum number correlations} \label{sec:x_kx_num_correlations}

Spatial inhomogeneous probing of ultracold atomic systems has been used to produce spatially structured density-density correlations \cite{Hauke2013a}, further enriched by the interplay between interaction and measurement dynamics  \cite{Pedersen2014a,Lee2014a,Mazzucchi2015a,Labouvie2015}. Similarly, the possibility of measurement induced momentum-momentum correlations is of interest \cite{Perrin2007a,Bucker2011a,Jaskula2012a,Finazzi2014a}.
In the following, we consider such correlations induced by continuous local probing with a spatially varying intensity profile and by stroboscopic probing with a planar-wavefront.

\begin{figure}[h]
  \centering
  \includegraphics[width=.48\textwidth]{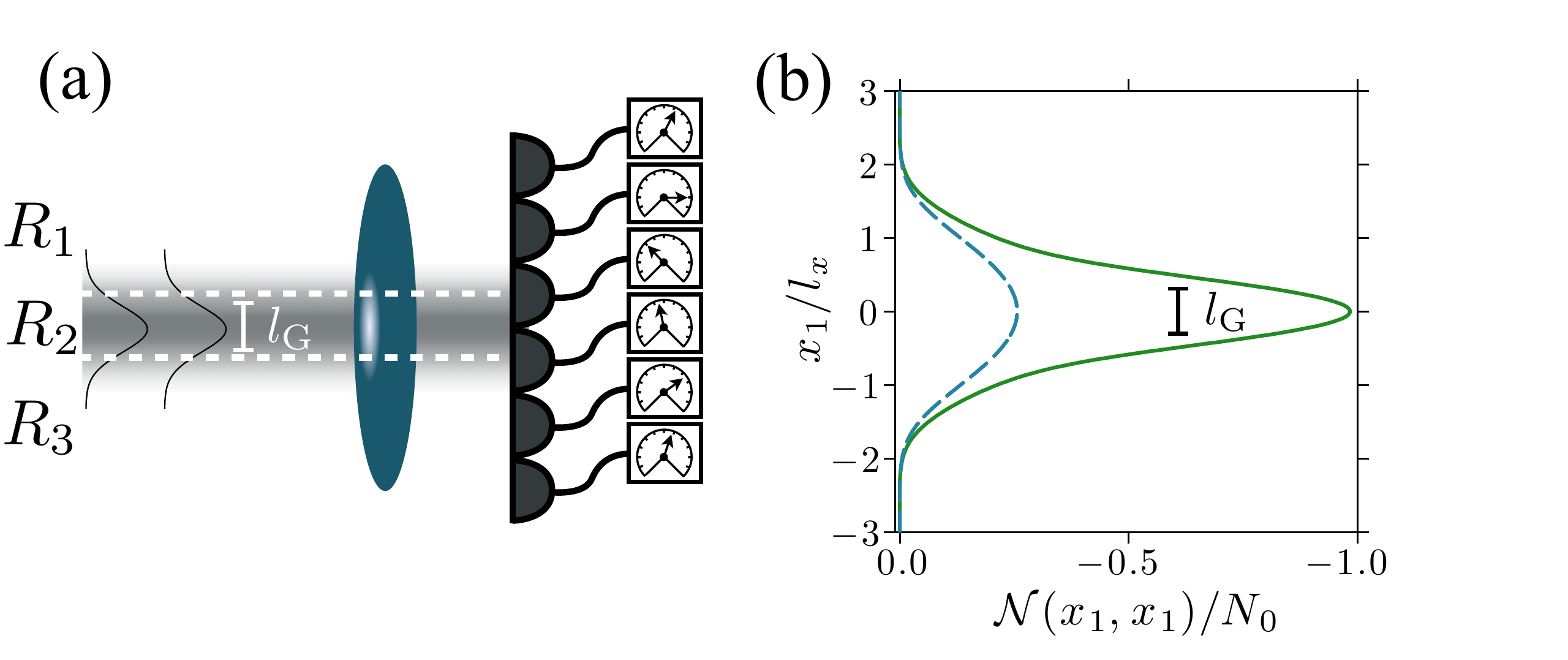}
\caption{(colour online) (a) The central region $R_2$ of width $l_\mathrm{G}$ is probed with a Gaussian beam of finite width $l_\mathrm{G}$. (b) Due to interactions, the local density fluctuations along the BEC axis are sub-Poissonian in the unprobed state (dashed-blue), and they are further suppressed by the probing in the region $R_2$ (solid-green). The solid-green line corresponds to the simulation of \figr{fig:LP_anal} with $l_\mathrm{G} = l_\mathrm{R}$. \label{fig:LP_cartoon}}
\end{figure}

We quantify the atomic density correlations by the covariance
\begin{equation}
\mathrm{covar}[\nh(x_1),\nh(x_2)] = \psi(x_1)\psi(x_2)\delta(x_1-x_2)+\mathcal{N}(x_1,x_2), \label{eq:den_fluct}
\end{equation}
where to second-order in atomic field quadratures,
\begin{multline}
\mathcal{N}(x_1,x_2) = -2\psi(x_1)\psi(x_2)\sum_{j} f_j^-(x_1) \Bigg\{ f_j^+(x_2) \\
-{}2\sum_{k} f_k^-(x_2) \mathrm{covar}[\hat{x}_j,\hat{x}_k]\Bigg\}. \nonumber
\end{multline}
The delta-function term in \eqr{eq:den_fluct} represents the Poissonian fluctuations, while non-Poissonian statistics and correlations between spatially separated regions lead to nonzero values of $\mathcal{N}(x_1,x_2)$.  Note that due to the atomic interactions, the Bogoliubov vacuum state already features sub-Poissonian fluctuations as indicated by the dashed-blue curve in \figr{fig:LP_cartoon}(b) \cite{Petrov2004a}.

\subsubsection{Continuous spatially inhomogeneous probing}
A spatially inhomogeneous intensity profile $u(x)$ of the light modifies our treatment of the light field through the replacement $\kappa_j(x)\rightarrow \kappa_j(x)\sqrt{u(x) l_\mathrm{L}}$. The mean field potential caused by the probe field yields a position dependent energy shift, but we assume for simplicity that the resulting potential can be canceled by adding a compensating potential.

\begin{figure}[h]
  \centering
  \includegraphics[width=.4\textwidth]{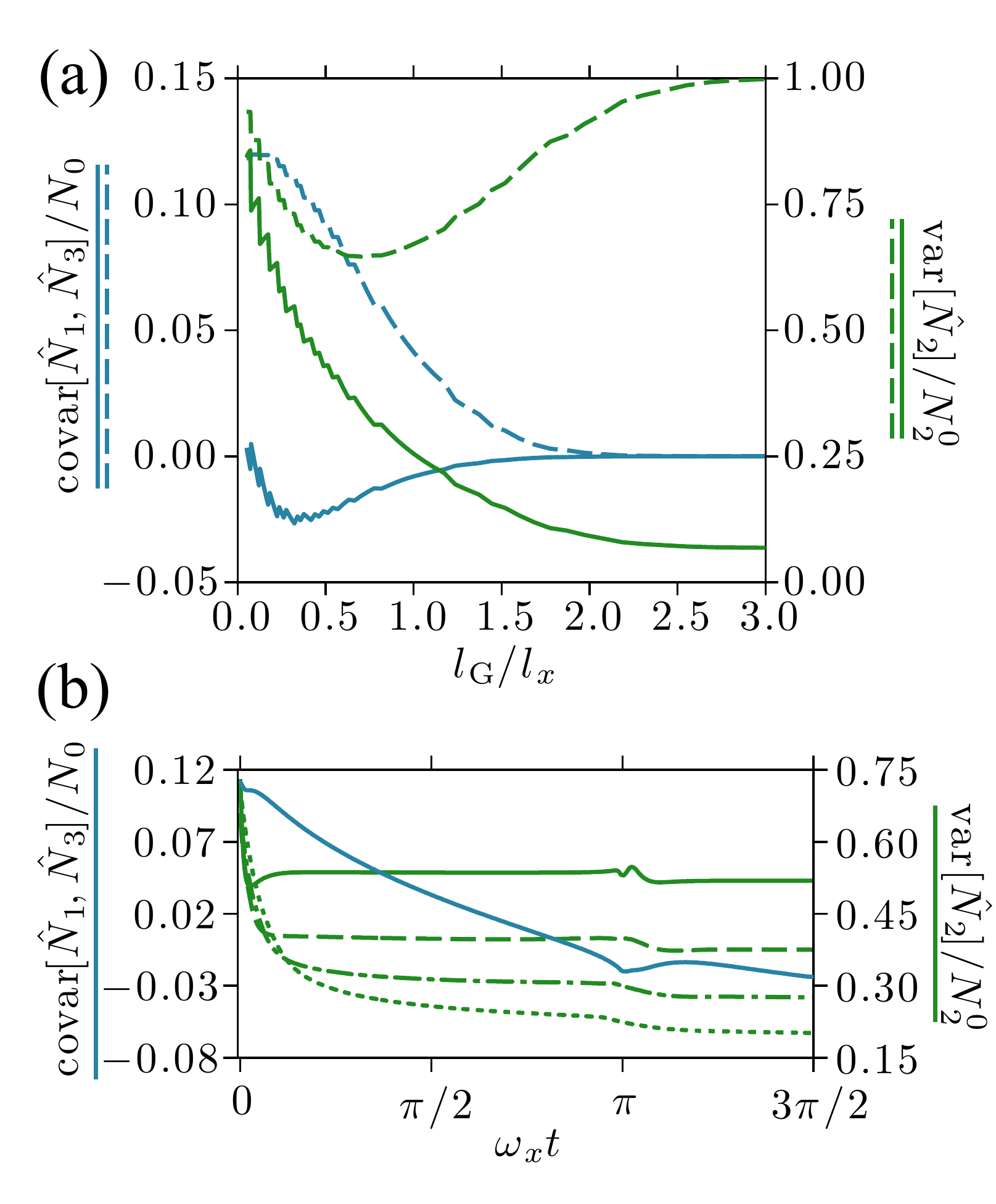}
\caption{(colour online) (a) The (blue) covariance $\mathrm{covar}[\hat{N}_1,\hat{N}_3]$ (normalized to $N_0$) between the number of atoms in the unprobed regions $R_1$ and $R_3$ and the variance $\mathrm{var}[\hat{N}_2]$ (green) of the number of atoms in region $R_2$ are shown, normalized by the BEC population in $R_2$, $N_2^0=\int_{R_2}\! \! dx n_0(x)$. The dashed curves show the results for the unprobed BEC. The solid curves show that after probing for a time $\omega_x t=3\pi/2$ with constant $\omega_x\bar{\kappa}^2_{00}=3/2$ and $l_\mathrm{D}=0.05 l_x$, $\mathrm{var}[\hat{N}_2]$ is reduced and the number fluctuations in regions $R_1$ and $R_3$ are reduced and can become anti-correlated. (b) The temporal evolution of the number fluctuations for $l_\mathrm{G}= l_\mathrm{R}$ (solid), $2l_\mathrm{R}$ (dashed), $3l_\mathrm{R}$ (dashed-dotted), and $4l_\mathrm{R}$ (dotted). The feature at $\omega_x t \sim \pi$ is explained in the text. \label{fig:LP_anal}}
\end{figure}

A Gaussian probe beam with width $l_\mathrm{G}$, \figr{fig:LP_cartoon}(a), leads to a reduction of the density fluctuations in the probed region, \figr{fig:LP_cartoon}(b). To discuss the interplay between the atomic dynamics, the measurement back action, and, the value of $l_\mathrm{G}$, we consider the three spatial regions indicated in \figr{fig:LP_cartoon}(a). First, we consider the fluctuations of total atom number $\hat{N}_2$ in the central region $R_2$ of width $l_\mathrm{G}$. In \figr{fig:LP_anal}(a), the dashed, green curve recovers the mean field Poissonian fluctuations when $R_2$ is wide and contains the total atom number, while the atomic interactions lead to sub-Poissonian fluctuations in any finite part of the BEC. Within very small regions, the population statistics again tend towards the Poissonian. The solid, green curve shows that probing suppresses the Poissonian number fluctuations of the whole BEC, and competes with the atomic exchange between the regions when the width of $R_2$ is decreased.
Eventually, the detector and diffraction resolution also become relevant.
For small $R_2$, higher-order modes are required to describe the number fluctuations and the role of the squeezed lower-order modes decreases. Eventually, the results become sensitive to the discrete structure of the modes, as seen by the discrete steps in \figr{fig:LP_anal} for lower $l_\mathrm{G}$ values.

We now consider the spatially separated regions $R_1$ and $R_3$.
If the system has a fixed total number of atoms, upon measuring $R_2$, squeezing $\hat{N}_2$, regions $R_1$ and $R_3$ can become anti-correlated (a squeezing of $\hat{N}_1+\hat{N}_3$).
In the Bogoliubov vacuum state, the Poissonian fluctuations of $\hat{N}_0$ cause $R_1$ and $R_3$ to share correlated fluctuations, \figr{fig:LP_anal}(a, dashed-blue).
Upon probing $R_2$, the coupling to the zeroth mode is held constant ($\omega_x\bar{\kappa}^2_{00}=3/2$), and consequently, $\hat{N}_0$ and $\hat{N}_2$ are squeezed leading to anticorrelated populations in $R_1$ and $R_3$, as shown by the solid, blue curve. This anticorrelation is limited as the fluctuations in $\hat{N}_1$ and $\hat{N}_3$ are individually squeezed.
For small $R_2$, the anticorrelations are lost, as the squeezing of $\hat{N}_0 \sim \hat{N}_1+\hat{N}_3$ is dynamically slowed by the strong coupling of the zeroth mode to many other modes.

For small $R_2$, the coupling of many modes induces a wave-packet in the fluctuations propagating towards the outer regions of the BEC and returning to the probed region with frequency $2\omega_x$ due to the linear spectrum of the higher-order modes populated by the wave packet. This causes the peak around $\omega_x t =\pi$ in the solid green and blue curves in \figr{fig:LP_anal}(b). As $l_\mathrm{G}$ increases, the feature is progressively lost with the decreased coupling to higher-order modes.

\begin{figure*}[t]
  \centering
  \includegraphics[width=.9\textwidth]{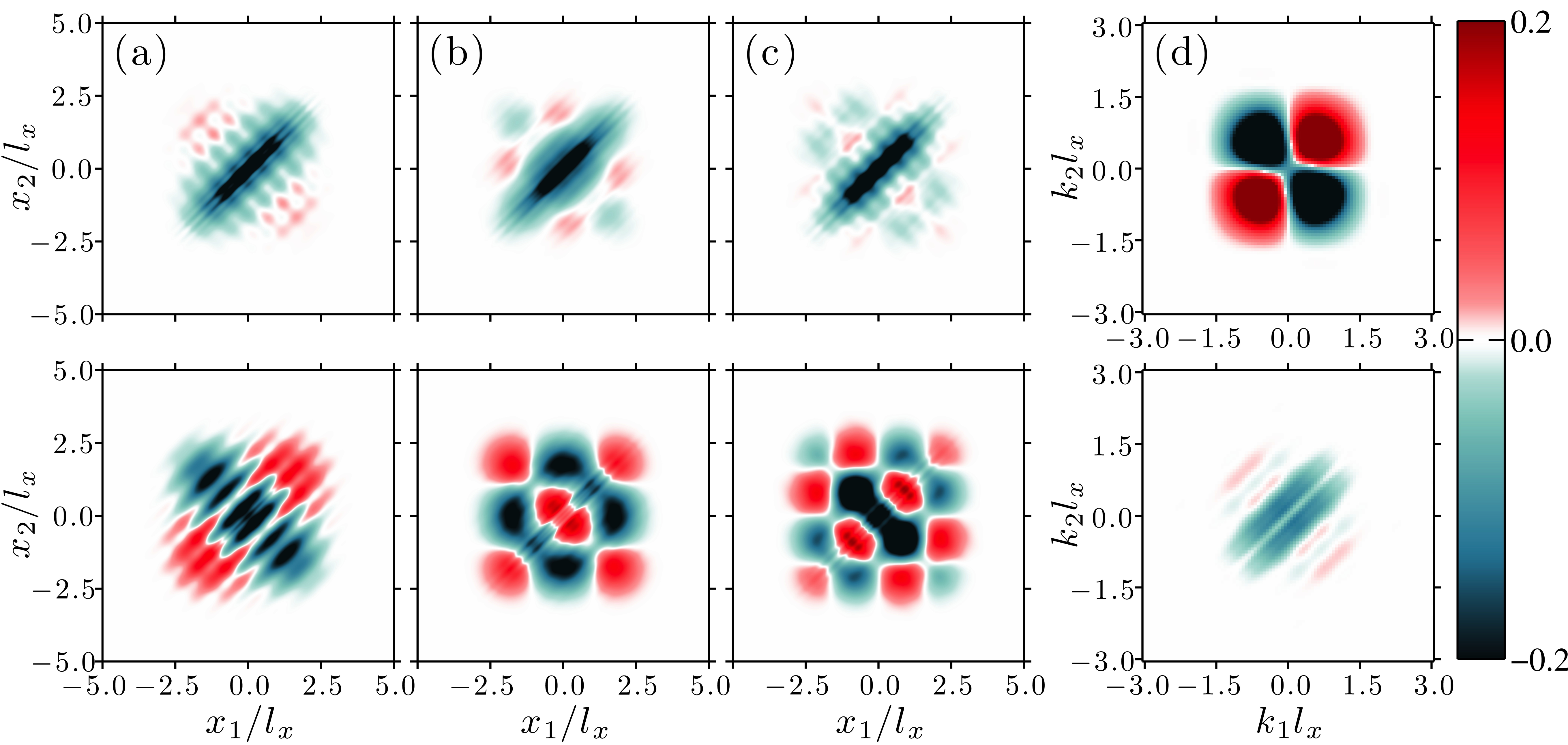}
\caption{(colour online) Controlled dynamical correlations. Modes $j=1$-$3$ (a-c) are squeezed with a train of $n$ pulses centred on times $t_l=[0,\pi/\omega_j,2\pi/\omega_j,\ldots,(n-1)\pi/\omega_j]$ with identical durations $\tau= (200 \omega_j)^{-1}$, $\kappa^2 = 100\omega_x/2\pi$, $l_\mathrm{D}=0.1 l_x$, and the total interrogation time of 100 trap cycles. The upper panels show $\mathcal{N}(x_1,x_2)$ when $\mathrm{var}[\hat{x}_j]=[0.0675,0.0587,0.0557]$ is minimal and for $j=[1,2,3]$, while the lower panels correspond to the maxima $\mathrm{var}[\hat{x}_j]=[3.7089,4.2692,4.5057]$. (d) shows the deviations from Poissonian momentum-momentum correlations for $\mathrm{covar}[\hat{m}(k_1),\hat{m}(k_2)]/5$ in the case of squeezing $j=1$. \label{fig:fluct_BIG}}
\end{figure*}

\subsubsection{Planar stroboscopic probing} \label{sec:strobo_den_fluct}

The spatially inhomogeneous probing of narrow subregions couple many atomic excitation modes, and upon suspension of the probing, the correlations will rapidly dephase.
Uniform stroboscopic probing, in contrast,  makes use of the natural evolution of the atomic system and yields mode selective squeezing that is persistent after preparation \cite{Wade2015a}.
The correlations associated with the modal structure of the targeted mode in position space for the first three density modes are shown in \figr{fig:fluct_BIG}(a-c). The upper(lower) panels show $\mathcal{N}(x_1,x_2)$ for times when maximal(minimal) squeezing occurs in the $\hat{x}$ quadrature of the targeted mode.
In addition, to the long wavelength modal structure of the targeted low energy modes, higher frequency spatial structure can be seen. In particular, in the cases of the $1$st and $3$rd modes. It turns out that the $1$st($3$rd) mode frequencies are almost commensurate with the $20$th($25$th) mode ($\omega_{20}/\omega_1=19.0005$, $\omega_{25}/\omega_3=8.9971$) leading to the small scale spatial correlations.

Figures \ref{fig:fluct_BIG}(d) show that structures due to the squeezing of the first mode in part (a) of the figure are accompanied by momentum density correlations $\mathrm{covar}[\hat{m}(k_1),\hat{m}(k_2)]$, where $\hat{m}(k) = \hat{\psi}^\dagger(k)\hat{\psi}(k)$ with $\hat{\psi}(k) = \int dx e^{ikx}\psihx/\sqrt{2\pi}$.

\section{Conclusions and Outlook} \label{sec:conclusions}

In this work we have demonstrated how dispersive optical probing allows control and preparation of quantum states of the collective vibrational motion of a BEC. Our Gaussian analysis permits the inclusion of the full quantum multimodal character of the problem, and, within the limitations of our expansion of the atomic field about the Gross-Pitaevskii mean field, we are able to take into account the atomic motion, interactions, and measurement back action in the full spatio-temporal evolution of the system.

The optical field is sensitive to the interference between the atomic mean field and the Bogoliubov modes, and three distinct types of resolution are identified associated with the (i) light-matter interaction, the (ii) light diffraction, and the (iii) detection of the light. The atomic interactions induce an irregular spectrum of density oscillations and results in our ability (i) to temporally resolve the atomic mode dynamics. 
The resolution associated with the diffraction of the light (ii) across the BEC restricts the addressing to a subset of the atomic modes.
Unresolved modes due to the spatial resolution of the detector (iii) leads to decoherence. By reading out only a particular distribution of pixels this decoherence may be engineered to affect one or a few modes. The measurement back action causes stochastic displacements of the atomic modes, and we have shown how to employ feedback to damp the displacements of specific modes. 

The results obtained here offer prospects for several interesting applications. Squeezing and entanglement are useful properties for precision sensing purposes, and the atomic system may find applications for sensing of field or inertial effects, if they couple to the atomic density. The controlled generation of mode-squeezing also makes this system a candidate for storage of quantum states of light \cite{Sherson2006a,Julsgaard2004a}, and the ability to address different collective modes paves the way for multimode quantum memory and repeater devices with a further potential to employ controlled interactions between the stored light modes.

Finally, the treatment here can be readily extended to a 2D pancake BEC, and the light mode diffraction across a 3D BEC could be incorporated in our treatment. Optical probing may here reveal the character of ordered classical and quantum phases, coherence dynamics and quantum transport, and, at the same time, contribute to the physical evolution of the system of interest. Analysis of the non-trivial interplay between interactions and measurements may well begin with systems for which our Gaussian analysis applies.

\begin{acknowledgments}
A.C.J.W.~and K.M.~acknowledge support from the Villum Foundation Center of Excellence, QUSCOPE, and the E.U.~Marie Curie program ITN-Coherence 265031.
J.F.S.~acknowledges support from Lundbeckfonden and the Danish Council for Independent Research.
A.C.J.W.~wishes to thank C.K.~Andersen and M.C.~Tichy for fruitful discussions and for feedback on the manuscript.
\end{acknowledgments}

\appendix

\section{Gaussian formalism} \label{app:gauss}

Gaussian states are characterised by only the first, $[\bnu]_j = \bracga{\hat{\bnu}_j}$, and second, $ [\bsig]_{jk}=\mathrm{cov}(\hat{\bnu}_j,\hat{\bnu}_k)$, moments of the canonical variables, $\hat{\bnu}=\bracb{\xh_1,\ph_1,\ldots,\xh_n,\ph_n}^T$,  which transform as $\bnu \rightarrow \bF \bnu + \bd$ and $\bsig \rightarrow \bF \bsig \bF^T$) under bilinear Hamiltonians and quadrature measurements \cite{Weedbrook2012a}. To describe the measurement transformation, the modes are separated into system and probe  modes
\eq{ \bsig=\bracb{\begin{array}{cc} \bA & \bC\\ \bC^T & \bB \end{array}},\qquad \bnu=\bracb{\begin{array}{c} \bR \\ \bQ \end{array}}. \nonumber}
$\bA$ ($\bB$) is the covariance matrix for the system (probe) modes, and $\bC$ is the covariance matrix between the two sets of modes, while $\bR$ ($\bQ$) are the first moments of the system (probe) modes. An ideal homodyne measurement of the momentum quadratures of the probe modes results in the transformation \cite{Fiurasek2002a,Giedke2002a,Eisert2002a}
\begin{subequations}
\label{APP:eq:update_eqs}
\begin{gather}
\bA \rightarrow \bA - \bC \bracb{\bLm \bB \bLm }^{MP} \bC^T, \\
\bR \rightarrow \bR - \bC\bracb{\bLm \bB \bLm}^{MP}\Delta \bQ,
\end{gather}
\end{subequations}
of the system modes, where $MP$ denotes the Moore-Penrose inverse, $\bLm =\diag([\blm^T,\ldots,\blm^T])$ with $\blm=[0,1]^T$, and for the $j^\mathrm{th}$ mode $[\Delta \bQ]_j= (p^\mu_j - \bracga{\ph_j})\blm$ with the measurement outcome $p^\mu_j$.

In the case of continuously monitored systems, the temporal limit of \eqrs{APP:eq:update_eqs} yields a differential matrix Riccati equation for the covariance matrix \cite{Stockton2004a,Madsen2004a}
\eq{\dot{\bA} =  \bE - \bD \bA  - \bA \bD^T - \bA \bM\bM^T \bA ,\nonumber}
from which the algebraic Riccati equation [$\dot{\bA} =\mathbf{0}$] follows, both being well known from classical optimal control \cite{Kwakernaak1972a}. The accompanying stochastic equation for the first moments is
\eq{d\bR =-\bD \bR  dt + \bA \bM \, d\bW ,\nonumber}
where the measurement result is now simulated with the Wiener increments $dW_j(t)$, $[d\bW]_j=dW_j(t)\blm$.
The matrix $\bE$ captures the environment back action, while the coherent evolution of the system enters through $\bD$. The measurement back action is represented by $\bA\bM$ and $\bA\bM\bM^T \bA$, depending on the measurement scheme [$\bM$] and the correlations [$\bA$].
As the unmeasured evolution ($\bM=\mathbf{0}$) corresponds to averaging over all measurement records \cite{Dalibard1992a}, we have the intimate link
$\bA_\mathrm{un}=\bA+\bA_\mathrm{ens}$
between the [un]measured covariance matrix [$\bA_\mathrm{un}$] $\bA$ and the covariance matrix $\bA_\mathrm{ens}$ of the trajectories $\bR$.

\bibliographystyle{apsrev4-1}

\end{document}